%% file: Report.tex
\pdfoutput=1
\documentclass[conference]{IEEEtran}

\input{Premables}

\begin{document}
\thispagestyle{plain}
\pagestyle{plain}
\pagenumbering{arabic}

\input{Chapters/00_Title}
\input{Chapters/00_Abstract}
\input{Chapters/01_Introduction}

\input{Chapters/02_LiteratureReview}
\input{Chapters/03_Methodology}

\input{Chapters/04_Results}
\input{Chapters/05_Conclusion}

\clearpage

\printbibliography

\end{document}

%% file: Premables.tex
\usepackage[utf8]{inputenc} 
\usepackage[english]{babel}
\usepackage[T1]{fontenc}
\usepackage{csquotes}
\usepackage{makecell}
\usepackage{cancel}
\usepackage{amsfonts}
\usepackage{amssymb}
\usepackage{graphicx}
\usepackage{subcaption}
\usepackage{textcomp}
\usepackage{float}
\usepackage{setspace, hyperref}
\hypersetup{colorlinks=false}
\usepackage{transparent}
\usepackage[final]{pdfpages}
\usepackage{multicol}
\usepackage{enumitem} 
\usepackage{parskip} 
\usepackage{amsmath}
\usepackage{amssymb}
\usepackage{latexsym}
\usepackage{mathtools}
\usepackage{physics}
\usepackage{amssymb}
\usepackage{comment}
\usepackage{arydshln}
\usepackage{pdflscape}
\usepackage{usebib}
\usepackage{tikz}
\usetikzlibrary{positioning}
\usepackage{comment}
\usepackage{booktabs}
\usepackage{xpatch}

\usepackage[style=ieee, backend=biber, uniquename=true, mincitenames=1, maxcitenames=2, citetracker=true]{biblatex}
\DeclareFieldFormat*{citetitle}{#1}

\usepackage{tabularx}
\newcolumntype{P}[1]{>{\hspace{0pt}}p{#1}}
\newcolumntype{K}[1]{>{\hspace{0pt}}X{#1}}
\newcolumntype{L}[1]{>{\raggedright\let\newline\\\arraybackslash\hspace{0pt}}m{#1}}
\newcolumntype{C}[1]{>{\centering\let\newline\\\arraybackslash\hspace{0pt}}m{#1}}
\newcolumntype{R}[1]{>{\raggedleft\let\newline\\\arraybackslash\hspace{0pt}}m{#1}}
\newcolumntype{C}{>{\centering\arraybackslash}X}

\addto\extrasenglish{}
\addto\extrasenglish{}
\addto\extrasenglish{}

\usepackage{titlesec}
\titleformat{\paragraph}[hang]{\normalfont\normalsize}{\theparagraph}{1em}{}
\titlespacing*{\subsubsection}{0pt}{1.5ex plus 1ex minus .2ex}{0.75em}
\titlespacing*{\paragraph}{0pt}{1.5ex plus 0.75ex minus .2ex}{1em}

\newcommand*\vcorrect{-1pt}
\usepackage{tikz,booktabs,array}
\makeatletter
\newcommand{\markZwicky}[1][]{\pgfutil@ifnextchar({\mark@Zwicky{#1}}{\mark@Zwicky{#1}()}}
\def\mark@Zwicky#1(#2)#3{%
   \tikz[every Zwicky picture,#1]{%
     \node[every Zwicky node,draw=none,inner sep=+\z@,outer sep=+\z@] {#3};
     \def\tikz@Mark@name{#2}%
     \ifx\tikz@Mark@name\pgfutil@empty\else
       \tikzset{every Zwicky node/.append style={name={#2}}}%
     \fi
     \node[every Zwicky node,overlay] {\phantom{#3}};
   }%
}
\newcommand{\tikzZwicky}[1][]{%
  \def\tikz@Zwicky@args{#1}%
  \let\tikz@Zwicky@list\pgfutil@gobble
  \let\tikz@Zwicky@first\pgfutil@empty
  \pgfutil@ifnextchar(\tikz@Zwicky@collect\tikz@Zwicky@finish
}
\def\tikz@Zwicky@collect(#1){%
  \ifx\tikz@Zwicky@first\pgfutil@empty
    \edef\tikz@Zwicky@first{#1}%
  \else
    \edef\tikz@Zwicky@list{\tikz@Zwicky@list,#1}%
  \fi
  \pgfutil@ifnextchar(\tikz@Zwicky@collect\tikz@Zwicky@finish
}
\def\tikz@Zwicky@finish{%
  \tikz[remember picture,overlay]
    \draw[every Zwicky connector,/expanded=\tikz@Zwicky@args]
      (\tikz@Zwicky@first) [/expanded={@Zwicky@list/.list={\tikz@Zwicky@list}}] [every Zwicky connect finish/.try];
}
\pgfkeys{/expanded/.code={\edef\pgfkeys@temp{{#1}}\expandafter\pgfkeysalso\pgfkeys@temp}}
\makeatother
\tikzset{
  @Zwicky@list/.style={insert path={to[every Zwicky connector how/.try] (#1)}},
  every Zwicky picture/.style={
    baseline,
    remember picture,
  },
  every Zwicky node/.style={
    remember picture,
    anchor=base,
    inner sep=+2pt
  },
  every Zwicky connector/.style={
    ultra thick,
    red!80!black,
    draw opacity=0.2,
    line cap=round,
    line join=round
  }
}

\makeatletter
\def\@IEEEsectpunct{.\ \,}
\def\paragraph{\@startsection{paragraph}{4}{\z@}{1.5ex plus 1.5ex minus 0.5ex}%
{0ex}{\itshape}}
\makeatother

%% file: Chapters/00_Title.tex
\begin{center}
	
\begin{spacing}{1}
{\LARGE \textbf{A Structured Survey of Quantum Computing for the Financial Industry}}
\end{spacing}
\bigskip
Franco D.\ Albareti\footnote{BME Inntech, Plaza de la Lealtad 1,
28014 Madrid, Phone: +34 91 709 56 59, Email: <\href{mailto:falbareti@grupobme.es}{falbareti@grupobme.es}>}, Thomas Ankenbrand\footnote{Hochschule Luzern, Institut f\"ur Finanzdienstleistungen Zug IFZ, Suurstoffi~1, 6343 Rotkreuz, Phone: +41 41 757 67 23, Email: <\href{mailto:thomas.ankenbrand@hslu.ch}{thomas.ankenbrand@hslu.ch}>}, Denis Bieri\footnote{{Hochschule Luzern, Institut f\"ur Finanzdienstleistungen Zug IFZ, Suurstoffi~1, 6343 Rotkreuz, Phone: +41 41 757 67 49, Email: <\href{mailto:denis.bieri@hslu.ch}{denis.bieri@hslu.ch}>}}, Esther H\"anggi\footnote{Hochschule Luzern, Lucerne School of Computer Science and Information Technology, Suurstoffi~1, 6343 Rotkreuz, Phone: +41 41 757 68 97, Email: <\href{mailto:esther.haenggi@hslu.ch}{esther.haenggi@hslu.ch}>}, Damian L\"otscher\footnote{Hochschule Luzern, Institut f\"ur Finanzdienstleistungen Zug IFZ, Suurstoffi~1, 6343 Rotkreuz, Phone: +41 41 757 67 94, Email: <\href{mailto:damian.loetscher@hslu.ch}{damian.loetscher@hslu.ch}>},  Stefan Stettler\footnote{Inventx AG, Haldenstrasse 23 8306 Br\"uttisellen, Phone: +41 81 287 16 72, Email: <\href{mailto:stefan.stettler@inventx.ch}{stefan.stettler@inventx.ch}>} and Marcel Sch\"ongens\footnote{SIX Financial Information AG, Pfingstweidstrasse 110, 8005 Z\"urich, Phone: +41 58 399 58 28, Email: <\href{mailto:marcel.schoengens@six-group.com}{marcel.schoengens@six-group.com}>}
\bigskip
\noindent
\\
\bigskip

\today

\end{center}

%% file: Chapters/00_Abstract.tex
\begin{abstract}
Quantum computers can solve specific problems that are not feasible on "classical" hardware. Harvesting the speed-up provided by quantum computers therefore has the potential to change any industry which uses computation, including finance. First quantum applications for the financial industry involving optimization, simulation, and machine learning problems have already been proposed and applied to use cases such as portfolio management, risk management, and pricing derivatives. This survey reviews platforms, algorithms, methodologies, and use cases of quantum computing for various applications in finance in a structured way. It is aimed at people working in the financial industry and serves to gain an overview of the current development and capabilities and understand the potential of quantum computing in the financial industry.
\end{abstract}

%% file: Chapters/01_Introduction.tex
\section{Introduction}\label{chap:Introduction}

Moore's Law states that the number of transistors in a microprocessor doubles approximately every two years. For decades, it has correctly predicted the increasing speed of computers and the constant reduction in the size of the chips that drives them. Despite this continuous increase in computing power, there are certain problems which are not solvable in practice today. Quantum computers --- computers which manipulate quantum information instead of ``traditional'' bits --- enable new approaches to certain complex problems. This could reduce the computing time required to solve previously untrackable problems to a manageable level.

A variety of applications of quantum computing for the financial industry have already been proposed (see e.g. \cite{ORUSOverview, egger2020quantum} for overviews). The propositions differ in the expected quantum speed-up, the necessary quantum resources, and the type of quantum computer required which makes it difficult to compare them directly. To this end, we propose a framework to analyze the use cases of quantum computing for the financial industry. It proceeds along layers, with each layer focusing on a different aspect.

This structured approach allows one to illustrate for which applications quantum computing is proposed in the financial industry and to what extent it can offer an advantage. Furthermore, it highlights how these applications are implemented and whether there are already results of corresponding experiments on quantum computing hardware. Quantum-inspired applications, such as tensor networks \cite{mugel2022dynamic}, are out-of-scope in this paper. These are interesting approaches, but they do not require quantum hardware.

The paper proceeds as follows: In \autoref{sec:LiteratureReview} we provide an overview of the current state of quantum computing. We also discuss the areas where quantum computing could potentially be applied in the financial industry and the implications of it. In \autoref{sec:Research_Methodology}, we propose a framework for a systematic analysis of proposals for the use of quantum computing in finance. We apply the framework to existing applications proposed in the literature in \autoref{sec:Results} and draw conclusions in \autoref{sec:Conclusion}.

%% file: Chapters/02_LiteratureReview.tex
\section{State of Quantum Computing in Finance}\label{sec:LiteratureReview}

\subsection{Current State of quantum computing research}

Quantum algorithms and hardware are an active area of research. In 2019, more than 100 academic groups and government-affiliate laboratories worldwide were researching how to design, build, and manipulate quantum bits (also called qubits) \cite{national2019quantum}. However, to build a quantum computer which can execute arbitrary quantum algorithms requires the ability to adaptively, reliably and precisely prepare, control and read out a large number of qubits.
The necessary requirements are described by the DiVincenzo criteria \cite{DiVincenzo_2000}:

\begin{itemize}
    \item A scalable physical system with well-characterized qubits
    \item The ability to initialize the state of the qubits to a simple fiducial state
    \item Long relevant decoherence times
    \item A "universal" set of quantum gates
    \item A qubit-specific measurement capability
\end{itemize}

\subsection{Hardware}

Various types of physical systems that could provide these properties are being researched. To build gate-based qubits the following five types are often considered~\cite{ladd2010quantum}: photons, trapped atoms, nuclear magnetic resonance, quantum dots, and superconductors. There is also a concept to use anyons (a type of quasiparticles) to build quantum computers \cite{Kitaev_2003}. Recent demonstrations of the power of quantum computational advantage over classical computers have been constructed using superconducting qubits \cite{arute2019quantum} and photonic qubits \cite{zhong2020quantum}. 
However, many of the physical systems on which current quantum computers are based cannot be scaled easily. 
For quantum computing to find a broader usage, more qubits are needed and the errors of actual hardware need to be reduced~\cite{WE2019}. Besides, the physical quantum computer architectures generally have constraints on qubit connectivity such as only nearest-neighbor couplings.
Additionally, one of the most critical aspects of quantum computers is the preservation of the quantum state of the qubits, i.e. the coherence of the qubits. This means that while qubits are under the control of the programmer, they must otherwise be isolated from the rest of the universe. This requires, for example, large cooling devices \cite{linke2017experimental}. 
In summary, the central challenge in building quantum computers is to maintain the simultaneous ability to control quantum systems, measure them, and at the same time preserve their strong isolation from uncontrolled parts of their environment \cite{ladd2010quantum}. All these points render it uncertain whether the current leading quantum technologies used to create early demonstration systems will be used to create future large-scale quantum computers.

\subsection{Simulators}

Since quantum computers are still at an early stage, quantum simulators have been developed. These are essentially software programs that run on classical computers and allow quantum programs to be executed and tested in an environment that predicts how qubits respond to operations. In this way, circuits of qubits can be tested to some degree without noise and with arbitrary connectivity. There are a variety of software libraries that can be used to simulate the behavior of qubits for different purposes.\footnote{There is an extensive list of tools for simulating the behavior of qubits, which can be found on \citeauthor{Quantiki2021List} \cite{Quantiki2021List}.}

In this context, it is important to discuss the exponential overhead associated with the simulation of quantum computers with classical computers. While a classical bit is limited to a value of either $0$ or $1$, a qubit could be in a combination of $0$ and $1$ states, a phenomenon which is known as quantum superposition. This superposition of states together with the postulates of quantum mechanics for composite systems implies that, while describing the state of a register with $n$ bits requires $n$ bits, describing the state of a quantum register with $n$ qubits requires $2^n$ complex numbers. It follows that simulating the behavior of a quantum computer on a classical computer is exponentially expensive in terms of the number of qubits~\cite{jones2019quest} and only feasible for small systems.

\subsection{Software}\label{subsec:Quantum_computer hard-_and_software}

Some of the hardware constraints, such as qubit connectivity or errors, can be circumvented by software. For example, swap operations \cite{nielsenchuang} can be used to apply gates between arbitrary qubits using the available (limited) connections. A swap gate simply interchanges the state of two qubits and can therefore be used to move the relevant logical qubits to two connected physical qubits where the desired gate can be applied. However, performing swaps increases the number of operations, which has two main drawbacks. First, it slows the algorithm down, and second, since all gates on a practical quantum computer are noisy, it leads to more overall noise which can even result in the algorithm becoming unusable in practice.

In a similar way, the challenge that quantum gates are noisy and quickly lead to calculations which are unreliable can be tackled using quantum error correction \cite{gottesman2009introduction}. This comes at the expense of needing even more qubits. Once quantum computers pass the size needed to implement error correction, they can correct the errors during a calculation, which will in turn allow even larger computations to be made. 

Further development of quantum software could reduce the hardware requirements of quantum computers and, in turn, lead to earlier applications of quantum computers in the financial services industry. While it can be a struggle to build the hardware that could support ambitious algorithms, the theoretical speed-up predicted for these algorithms is also a reminder of the potential of quantum computing. Algorithm research and hardware development therefore incentivize each other~\cite{schulte2019ai}.  Experimental progress is necessary to unlock the full potential of quantum computing. However, it is possible that noisy intermediate-scale quantum computers will find interesting applications far before general-purpose, fault-tolerant quantum computing is available~\cite{Orus2019}. 

Another factor that needs to be mentioned when discussing quantum computer development is the availability of open source software projects that lower the barrier to learn quantum computing and thus accelerate its use. Understanding, creating, and executing complicated mathematical models on quantum computer hardware has become easier with the development of open source software projects~\cite{Fingerhuth_2018}. The next ten years are expected to be the decade of quantum systems and the emergence of a real hardware ecosystem that will provide the foundation for improving quantum computing~\cite{IBMFlight}.

\begin{figure*}[ht]
	\centering
	\includegraphics[trim=1.5cm 1cm 9.5cm 9.5cm,clip,width=1\linewidth]{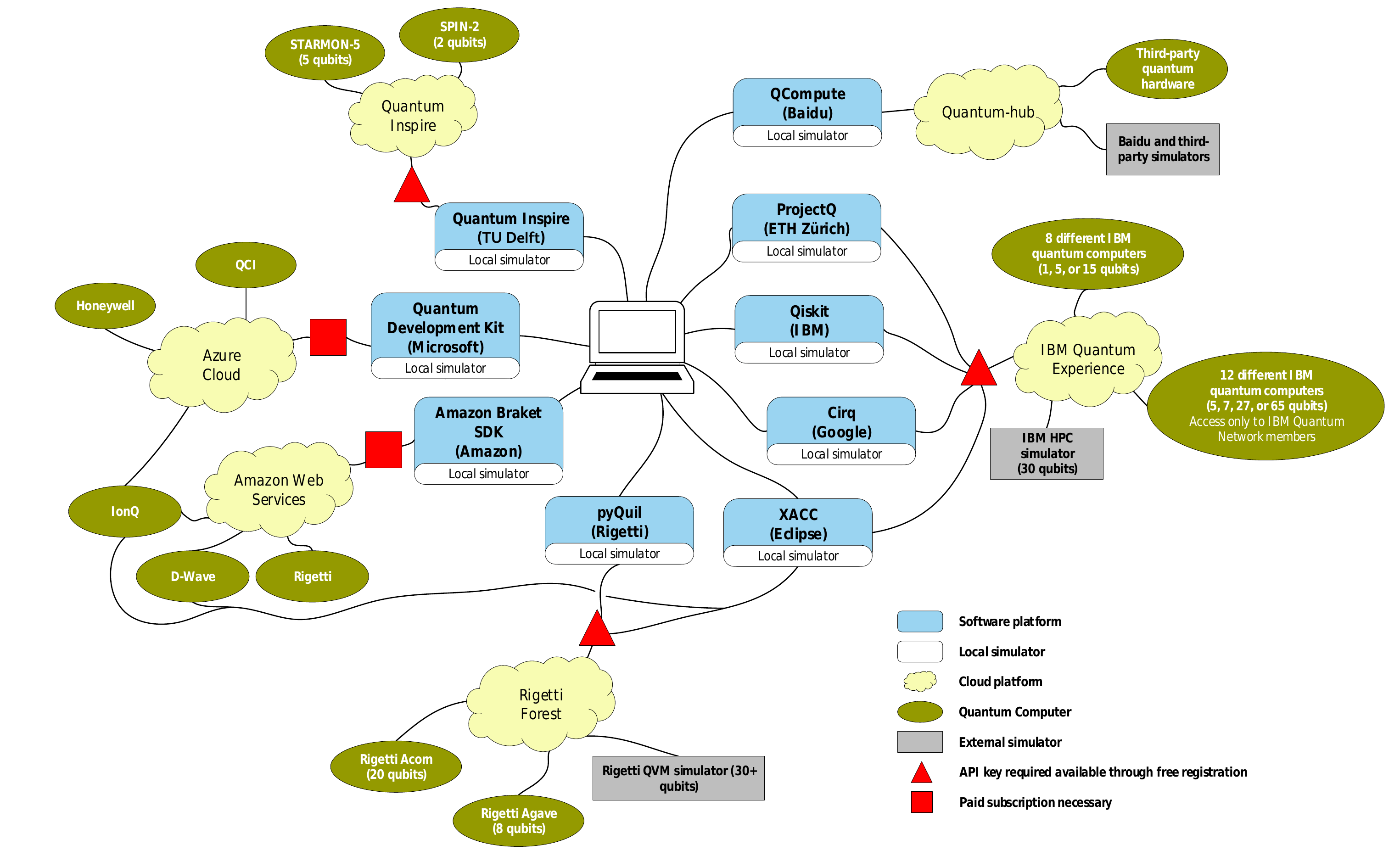}
	\caption{The nodes in light blue show software platforms that can be installed on the user’s personal computer. White nodes show simulators run locally, i.e. on the user’s local device. In this case, the number of qubits that can be simulated is dependent on the performance of the user's local computer. The quantum computing company resources are shown in clouds. The red triangles and squares represent the requirements to get access to the corresponding resource. The quantum simulators and usable quantum computers provided are shown in dark green. For example, to connect to IBM’s cloud (light green) and use one of their quantum computers, one needs to download and install the quantum programming software QISKit locally, register on IBM’s website to get an API key, and then request access to the quantum device.\protect\footnotemark The number of qubits of the quantum computers are written in brackets, where the corresponding information is publicly available.}
	\label{fig:Landscape}
\end{figure*}

Although more qubits are needed to solve more complex problems and quantum coherence (the time during which a quantum state is preserved) and topology are limited (e.g. logical operations cannot be applied to every qubit), noisy intermediate-scale quantum computers are already available to the public. Furthermore, there is a range of open source software and quantum computer platforms publicly accessible to connect a personal computer to a quantum computer over the cloud; see, e.g., ~\cite{Fingerhuth_2018} for a review. To run an algorithm on a real quantum computer requires the usage of a software with a quantum computer back end. In addition, a full-stack library allowing for the compilation/embedding, simulation, and execution of quantum instructions on a quantum computer is needed to implement a specific use case. \autoref{fig:Landscape} shows the software platforms as of the end of January 2021, using a similar representation as~\cite{LaRose_2019}. 

\footnotetext{Qiskit is an open source software development kit for working with quantum computers at the level of pulses, circuits, and application modules. For further information, see \url{https://qiskit.org/}.}

\subsection{Fields of quantum computing in finance}\label{subsec:Fields_of_quantum_computing_in_finance}

Although there are still hurdles to overcome in terms of the development of quantum computer hardware and software, there are already multiple publications that suggest potential applications of quantum computing in the financial industry. In \cite{ORUSOverview} the following three main fields of application are discussed:

\begin{itemize}
	\item Optimization
	\item Machine learning
	\item Monte Carlo methods
\end{itemize}

In \cite{egger2020quantum} the same categorization is employed, but the more general term "Simulation" instead of "Monte Carlo" is used. In this paper, the authors also provide examples for concrete use cases. Furthermore, they point to three areas of financial services, where problems currently arise that pose a challenge for classical computers. These areas include asset management, investment banking, and retail and corporate banking. There are also numerous other potential future fields of application related to quantum technology such as quantum simulation, quantum money and quantum cryptography \cite{ORUSOverview}. In \cite{IBM2019}, the authors see potential applications in features selection, product recommendations, arbitrage, portfolio management, derivative pricing, risk analysis, credit scoring, fraud detection, anti-money laundering, and forecasting financial crashes. 

Hence, there is a wide variety of proposed applications for quantum computing in the financial industry, even though the technology is still in its infancy. This underscores that quantum computing could have a significant impact on the financial industry in the future.
\enlargethispage{-3\baselineskip}

\subsection{Impact of quantum computing on the financial industry}

The impact of quantum computing on society is discussed in \cite{Wolf}. The author writes that many areas such as secure digital communication, money transfer, and signature of digital documents are based on concepts of cryptography. Using a quantum computer to break cryptography would have a major impact on our economy and society. It is easy to follow his conclusion that a world without reliable electronic payments and banking transactions would grind to a halt. As for faster search/optimization and simulation, from a utilitarian perspective, increased efficiency in areas such as planning, resource allocation, machine learning, and development of new drugs and materials can benefit society as a whole.

The literature on the impacts of quantum computers on the financial industry or an estimate of the potential monetary benefits is scarce. It was suggested in \cite{IBM2019} that applying quantum computing in finance results in enhancing investment gains, reducing capital requirements, creating new investment opportunities, and improving risk management and compliance.  It was stated in \cite{Atos} that application of quantum computing could lead to an increase in cyber-security, accelerate high-frequency trading, or lead to greater accuracy in the simulation of customers' purchasing preferences, resulting in an improved data-driven customer relation management.

The authors in \cite{ORUSOverview} write that the race in the field of quantum computing is largely motivated by the disruption the technology is expected to bring, resulting in a complete transformation of the financial services industry. However, the authors do not elaborate on the expected degree of disruption or impact. The high expectations might be explained by the theoretical speed-up compared to classical computers. However, it is still very difficult to estimate when the technology will be available to exploit the potential of these algorithms. For instance, in \cite{Kumar} it is suggested that we stand at least five years away from quantum computing significantly impacting the financial services landscape. This is in line with the authors of \cite{IBMFlight} who state that widely-adopted commercial applications may remain several years away. Also, there are no estimates as to how expensive maintaining the infrastructure and thus running a quantum computer will be.

%% file: Chapters/03_Methodology.tex
\section{Research Methodology}\label{sec:Research_Methodology}

In this section, we propose a framework to systematically review the quantum computing applications proposed in the literature for the financial industry. In the first subsection, we discuss the selection of relevant literature. In the second subsection, we propose the framework that is applied to the selected literature and the use cases described in it.

\subsection{Literature Selection}\label{sec:search_for_existing_literature}

As shown in \autoref{sec:LiteratureReview}, various potential areas for quantum computing in the financial industry have already been identified. To analyze specific use cases in the financial industry in more detail, we first search for relevant literature on the main source: Google Scholar. With this search, a broad selection of academic literature on the topic can be covered and thus, the existing research be identified. For the search itself, papers are excluded that provide panel introductions; papers that are not available in English nor German; and teaching cases and pedagogical research papers. In a next step, the selection of the final sample is performed. Only papers that describe a concrete application of quantum computing for the financial industry or its implementation on a corresponding device are considered. In addition, a requirement is that the paper must be peer-reviewed and published by 2021 at the latest in order to be considered. In total, 13 papers that meet all requirements are identified and are analyzed in the next subsection.

\subsection{Layered Framework Analysis}

\begin{figure}[h]
	\centering
	\begin{tikzpicture}[
    squarednode/.style={rectangle, draw=black!60, fill=black!10, very thick, minimum size=10mm, minimum width=\linewidth},
    ]
    \node[squarednode] (usecase) {Use Case};
    \node[squarednode] (methodology) [below=0.25cm of usecase] {Methodology};
    \node[squarednode] (quantumalgorithm) [below=0.25cm of methodology] {Quantum Algorithm};
    \node[squarednode] (quantumhardware) [below=0.25cm of quantumalgorithm] {Quantum Hardware};
    \end{tikzpicture}
	\caption{Main layers for the use of quantum computing in finance.}
	\label{fig:Main_Layers}
\end{figure}
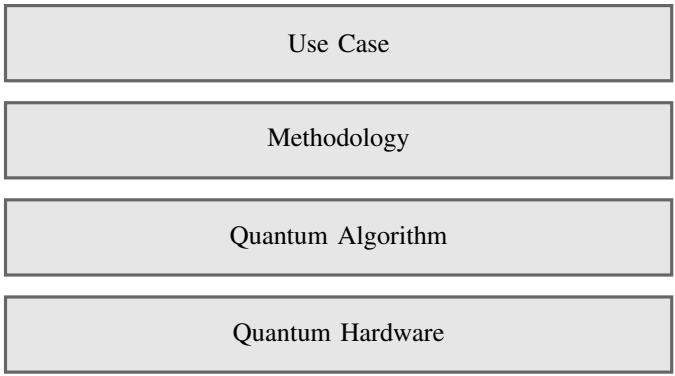

The selected literature proposes a variety of applications of quantum computers for the financial industry. However, the proposed implementations and expected advantages of these applications differ in many aspects. Due to these differences, it is difficult to compare them directly. We therefore develop a framework to compare the applications of quantum computing for the financial industry. We neglect operational aspects such as manufacturer-specific Application Programming Interfaces (API) and software needed for the interaction with the quantum computer, but instead focus on a conceptional point of view. The framework divides each proposal into four different layers, as shown in \autoref{fig:Main_Layers}. While the upper layers are closely tied to the task from the financial industry which shall be solved, the lower layers are related to the technical aspects of implementation of quantum computers.

The top layer describes the use case in finance and the second layer contains the corresponding methodology like optimization, machine learning, and Monte Carlo simulation. Most applications described in the literature are based on a small set of existing quantum algorithms and use its speed-up to improve the results (speed, precision etc.) for the use case in the financial industry. Therefore, the third layer describes the quantum algorithm used. The last layer is about the quantum hardware for which an implementation has already been developed. It also includes an assessment of whether the algorithm has already been run on appropriate hardware and, if so, with what result. 

Note that the four layers are not completely independent: the same way that the methodologies must be selected according to the use case, there is not one quantum algorithm which fits all methodologies. Some of the algorithms are 
more suitable for near-time prototype quantum computers and others require the implementation of larger quantum computers, but come with the advantage of theoretically proven speed-ups (see also \cite{bouland2020prospects}). However, this separation into layers allows us to see the differences between the various use cases of quantum computing in finance.

The specific focus of each of the four layers is described in more detail in the following paragraphs. For each layer, we include examples to illustrate its meaning.

\subsubsection{Use Case}\label{subsubsec:Met_Use_Cases}

The use case layer describes the specific problem for which quantum computing can be useful in finance and what the concrete benefits are. The benefit should result, for example, in lower costs, more revenue, or lower risk, and ultimately in higher profitability of existing business areas or the development of new business opportunities. This can be achieved by methods such as calculating the price of a financial product more quickly, which can open up an arbitrage opportunity. A benefit can also result if reputational and financial risks in connection with, e.g., fraud can be minimized.

\paragraph{Portfolio Optimization}
An important use case is portfolio optimization, where the aim is to optimize asset allocation optimizing a certain objective function, e.g. maximizing the Sharpe ratio \cite{Sharpe1994Sharpe}. In \cite{bouland2020prospects} it is emphasized in that although portfolio optimization in its simplest form is relatively easy to solve, as it can be reduced to solving a single system of linear equations, the computational difficulty of this problem increases when more realistic constraints are added to the problem. 

\paragraph{Transaction Settlement}

Transaction clearing is one further use case of quantum computing. The clearing house's goal is usually to settle as many transactions as possible or to maximize the total value of settled transactions, but this is a difficult optimization problem due to a combination of legal frameworks and additional options. However, if more transactions could be cleared by using a quantum computer, this increases the efficiency of markets (for an overview see \cite{gedin2020securities}). 

\paragraph{Estimating Risk Measures}
A quantum computer could also be used to calculate risk metrics such as Value-at-Risk (VaR), Conditional VaR (CVaR) or credit risk. The benefit of using a quantum computer is that these measures can be calculated faster or more accurately. This brings great advantages for financial institutions, such as optimizing risk-bearing capital or higher profits through lower credit defaults. 

\paragraph{Pricing Financial Derivatives}
Another application is in the pricing of financial derivatives, which can be complicated functions of market prices. A significant short-term advantage for first movers is to price actively traded derivatives and bonds faster than other market participants. In addition, pricing bonds and derivatives using a quantum computer instead of classical Monte Carlo simulation could contribute to a more accurate valuation, especially for less liquid traded derivatives and bonds with complex payout structures and contract terms. The reason for this is that more price-driving factors could be taken into account compared to simulations with classical computers. An advantage arises especially at the stage when only a limited number of players have exclusive access to faster calculation capabilities. 

\paragraph{Predicting Financial Crashes}
Quantum computers could make better predictions of financial crises. The topic is highly relevant from both micro- and macroeconomic perspectives and is a central issue in the study of macroeconomic stability. The predictability of financial crises is widely considered to be low and very complex (for an overview see \cite{greenwood2020predictable}). The many financial actors and the hardly-predictable market psychology (behavioral finance) are only two factors contributing to this fact. Quantum computing could help to calculate complex models that take into account many actors with their complex decision-making processes and their interrelationships.

\subsubsection{Methodology}

The literature discussed in \autoref{sec:LiteratureReview} mainly refers to three areas in which quantum computing can be useful for the financial industry. In our framework, we classify the relevant literature identified (\autoref{sec:search_for_existing_literature}) according to the same areas.

\paragraph{Optimization} Optimization problems are the core of many financial problems for portfolio optimization or rebalancing. These kinds of problems are thus seen as potential use cases of quantum computing.

\paragraph{Monte Carlo (simulation)} Monte Carlo (simulation) can be used for modeling
complex systems such as the value of an asset. Using a quantum computer instead of
traditional Monte Carlo (simulation) could contribute to a more accurate/quicker solution of complex problems.

\paragraph{Machine learning} The use of machine learning concepts, which are already used widely in the financial sector, are expected to be used even more in the future and represent another promising application field for quantum computing.

\subsubsection{Quantum Algorithms}\label{subsubsec:Met_Quantum_Algorithms}

From a theoretical point of view, it can be shown that a quantum computer can compute anything a classical computer can compute~\cite{nielsenchuang}. Of special interest are, of course, quantum algorithms which give an advantage over classical algorithms. Which problems can or cannot be solved efficiently on a quantum computer is still an open question in theoretical computer science, just as it is for classical computers. Research in quantum computing has also spurred progress in classical algorithms, for instance as described in~\cite{gilyen2018}, \cite{tang19} and \cite{gilyen2020}. Therefore, for most problems it remains open whether quantum computers have an advantage over classical computers.

From a more practical point of view, a second question arises: for which problems are current and near-term quantum computers useful? Such an analysis must take into account the hardware constraints of \emph{real} quantum computers, e.g.\ time it takes to implement specific physical transformations (i.e.\ `clock speed' of the quantum computer), their size (i.e.\ number of qubits), noise level, transformation of initial inputs from classical computers to quantum hardware, etc. This question has also lead to the development of algorithms which are better suited to run on present hardware, such as hybrid quantum-classical algorithms for optimization problems. Though selectively there are results showing supremacy of quantum algorithms, to the best of our knowledge there is no rigorous end-to-end consideration to paint the full picture.

Below, we describe the quantum computing algorithms often discussed in the literature in connection with quantum computing in finance. The reader is referred to Ref. \cite{abhijith2018quantum} or \cite{Jodan2021Quantum} for a concise and more comprehensive list of quantum algorithms. In this paper, we distinguish between the following algorithms:

\paragraph{Searching and counting}\label{par:Grovers_search_algorithm}

\citeauthor{grover1996fast} \cite{grover1996fast} presents a search algorithm that needs quadratically fewer queries in an unstructured search compared to any classical algorithm and illustrates it with searching a specific phone number in a phone book. More precisely, a phone book with $n$ unsorted entries could be searched in $O(\sqrt{n})$ steps on a quantum computer as opposed to $O(n)$ steps on a classical computer. 

\citeauthor{Brassard2000Quantum} \cite{Brassard2000Quantum} present generalizations of Grover’s search algorithm, called "Quantum Amplitude Estimation (QAE)" and "approximate counting", which can be used to estimate the number of items in a set fulfilling a certain condition and are core subroutines in quantum computation for various applications. The traditional approach for amplitude estimation is to use the Quantum Phase Estimation (QPA) algorithm, which consists of many controlled amplification operations followed by a quantum Fourier transform \cite{nielsenchuang,Suzuki_2020}. 

Recently, several improved variants of Quantum Amplitude Estimation have been introduced. One variant does not rely on QPE but is only based on Grover's algorithm, which reduces the required number of qubits and gates~\cite{Grinko_2021}. Another approach~\cite{Suzuki_2020} allows the quantum amplitude estimation algorithm to be implemented without the use of expensive controlled operations.

Approximate counting is especially useful to estimate statistical properties of a sample such as it is `traditionally' often done by Monte Carlo methods. 

\paragraph{Solving linear equation systems}

\citeauthor{HarrowHassidimLloyd09} \cite{HarrowHassidimLloyd09} propose an algorithm for solving systems of linear equations. More precisely, the algorithm calculates the approximate result of a function of the solution $x$ to the linear equation $Ax=b$. However, there are various restrictions on the exact type of problems this algorithm can solve~\cite{aaronson15,bouland2020prospects} which might explain why there are only very few papers using this method in the context of finance~\cite{Rebentrost_2018}. 

Nevertheless, solving linear equations is a fundamental building block for many algorithms, including machine learning algorithms. The Harrow-Hassidim-Lloyd (HHL) algorithm has lead to several propositions for general-purpose fast machine learning using quantum algorithms (e.g.,~\citeauthor{KerenidisPrakash16} \cite{KerenidisPrakash16},~\citeauthor{KerenidisLandmanPrakash19} \cite{KerenidisLandmanPrakash19},~\citeauthor{Allcocketal20} \cite{Allcocketal20} or~\citeauthor{Liu_2018} \cite{Liu_2018}), including a quantum version of principal component analysis~\cite{Lloyd_2014} (qPCA). 

The speed-up achieved by the quantum machine learning algorithms depends on the type of problem to be solved, but may be polynomial or even superpolynomial~\cite{Jodan2021Quantum}.\footnote{The proposed fast algorithms for recommendation systems running on quantum computers have lead researchers to ask the question whether there could be better classical algorithms for this problem. This has lead to the discovery of classical algorithms~(\cite{tang19}) which are much faster than the previously known ones.} To apply any of the above quantum machine learning algorithms requires, however, for the problem to be formulated in a very specific form. Bringing a realistic problem into this form can be challenging and may ruin the speed-up which could be gained from the quantum machine learning algorithm~\cite{aaronson15,Tang21}.

\paragraph{Optimization}

Improvements to solve optimization problems such as semi-definite programs via a quantum version of the interior point method (QIPM) have been proposed by ~\cite{brandao2017quantum,brandao2019quantum,kerenidis2018quantum}. These algorithms are based on Gibbs sampling~\cite{Poulin_2009} and may reach a quadratic speed-up~\cite{Jodan2021Quantum}. Due to the generality of semi-definite programming problems, which include e.g.~second order cone programs (SOCPs) or linear programs (LPs) possible applications of improvement are vast. 

\paragraph{Approximate Optimization}

In our analysis, each layer represented in \autoref{fig:Main_Layers} is considered in isolation. However, a completely isolated view is only partially possible for quantum algorithms. While the algorithms discussed above are suitable for gate-based quantum computations, the Ising model forms the basis for quantum annealing (see Paragraph~\ref{par:annealers}). This model can be used to solve combinatorial constraint satisfaction problems~\cite{Farhi2000,farhi2001}. 

For many combinatorial optimization problems it is even hard to find an approximate solution which is close to the optimum~\cite{sahni1976}. On the other hand, there often exist algorithms which are known to find reasonably good approximations. The Quantum Approximate Optimization Algorithms~\cite{Farhi2000,farhi2001,farhi2015quantum} serve to find approximate solutions faster than the currently known classical algorithm by exploiting quantum fluctuations \cite{das2005quantum}, i.e., tunnelling effects between local minima. Amongst the most widely used algorithms for this type of procedure are the Variational Quantum Eigensolver (VQE)~\cite{Peruzzo2014} and the Quantum Approximate Optimization Algorithm (QAOA)~\cite{farhi2014quantum}. Both of these algorithms are classical-quantum hybrid algorithms, i.e., they combine the use of a classical computer with a (smaller) quantum computer which can further improve the applicability of these algorithms on current quantum computing devices~\cite{liu2021layer}. 

A special form of combinatorial optimization problems are so-called Quadratic Unconstrained Binary Optimization (QUBO) problems. They unify a variety of combinatorial optimization problems that are often encountered in the financial industry. QUBO can be mapped to the Ising model, which allows the application of quantum algorithms such as VQE or QAOA~\cite{lucas2014}. 

Note that in the meantime there have been improvements in \emph{classical} approximation techniques for the constraint satisfaction problems where QAOA are typically used~\cite{barak2015}. The exact quantum advantage for these problems therefore remains an open research question. 

\paragraph{Further algorithms}

The above discussion includes only a small selection of all proposed quantum algorithms. A current list with many more algorithms can be found on~\cite{Jodan2021Quantum,Montanaro2016}.

We have not discussed any of the algorithms for the hidden subgroup problem, which includes Simon's algorithm~\cite{Simon1997} and Shor's algorithm~\cite{Shor_1997}. Shor's algorithm is among the best-studied quantum algorithms, since it allows for fast integer factorization and taking discrete logarithms --- thereby threatening many of the currently used crypto systems~\cite{rsa,dh}. We are not aware of any results which directly apply these algorithms in finance.

\subsubsection{Quantum Hardware}

The quantum hardware being considered actually for problems in finance can be classified into gate-based quantum computers and quantum annealers. 

\paragraph{Gate-based quantum computers}

Most of the algorithms discussed above are intended for gate-based quantum computers. A gate-based quantum computer system relies on the construction of reliable qubits on which a sequence of elementary quantum logical gates can be executed after each other. This is similar to the operations of a classical computer where a computation consists of applying elementary operations such as NOR, AND, OR or NOT. The elementary gates can be assembled into an arbitrary sequence to execute a quantum algorithm. The input is the initial state of the quantum register, comprising of all the qubits of the quantum computer. At the end the final state of the quantum register, i.e. the result of the computation, is measured to obtain a classical result~\cite{nielsenchuang}. 

This class of quantum computers is difficult to scale in practice \cite{Preskill2018quantumcomputingin}. Current gate-based quantum computers still have a very limited number of qubits and are error-prone. While error correction is possible \cite{shor1995error}, this introduces a large overhead (see, e.g., \cite{roffe2019quantum} for an introduction). Nevertheless, most of the quantum computers depicted in Figure~\ref{fig:Landscape} are gate-based and the recent experiments showing an advantage were achieved on this type of quantum computer. 

\paragraph{Quantum annealers}\label{par:annealers}
 
 Quantum annealing is based on a general result of quantum mechanics known as the adiabatic theorem (quantum annealing is also known as adiabatic quantum computation ~\cite{albash2018adiabatic}). The idea behind this theorem is that, when starting from a physical system in its ground state (the state with minimum energy) and slowly (!) modifying it, the system will remain in the ground state. The ground state of the initial system could be easily prepared while the ground state of the final system is the solution that is looked for. This is due to the fact that the final physical system has been tuned in such a way its ground state represents a solution to an actual problem of interest. Quantum annealing is closely related with, but not limited to, optimization problems.
 
 Even though the way computation is represented in adiabatic quantum computing seems very different from gate-based quantum computing, the two can be shown to be equivalent from a theoretical perspective~\cite{aharonov2007}. Adiabatic quantum computing is hoped to be more easily scalable and less sensitive to errors than gate-based quantum computers.

%% file: Chapters/04_Results.tex
\section{Results}\label{sec:Results}

In this section we present the 13 selected papers and the results of the analysis according to the layer framework. In the first subsection we give an overview of the properties in the different layers. In the following subsections, we then discuss the results of the corresponding papers in detail.

\subsection{Overview}\label{subsec:Overview}

The papers are presented in the morphological box in \autoref{fig:morphologicalbox}. The attributes of the different layers are shown horizontally. Each paper is represented by a blue line. If identical properties are present for several papers, the corresponding line depicted is superimposed and therefore appears darker. In addition, all papers are shown in tabular form in \autoref{tab:Table}. This helps to get a rough overview. In the following (\autoref{subsec:Portfolio_optimization} to \ref{subsec:Derivative_Pricing}), we then discuss the papers in detail.\input{Figures/Zwicky-Box}

\subsubsection{Use Cases}\label{subsubsec:Use_Cases}

The papers propose use cases for quantum computing in finance in five different areas. These include portfolio optimization, transaction settlement, financial crash prediction, derivatives pricing and risk measure estimation. The breakdown of the 13 papers across these areas is shown in \autoref{fig:Use_Cases}. It is noticeable that five papers focus on the calculation of optimal portfolios with the help of a quantum computer. Furthermore, there is one cluster with three papers discussing the pricing derivatives with quantum computers.\input{Figures/Use_Cases}

\subsubsection{Methodologies}\label{subsubsec:Methodologies}

As shown in \autoref{fig:Methodologies}, eight of the 13 papers considered relate to optimization, six to Monte Carlo simulation and none to machine learning. The lack of papers focusing on machine learning is surprising as such methods are already applied today using traditional computers. The reason for the different number of papers in the three different methodologies can be explained by the hardware requirements \cite{bouland2020prospects}. They described that it has been shown that full-scale fault-tolerant (gate-based) quantum computers provide a significant speed-up by drastically reducing the number of samples required compared to classical Monte Carlo methods \cite{HEINRICH20021, Montanaro_2015}. The main limitation for optimization algorithms (for gate-based quantum computer) is their dependence on quantum linear system solvers \cite{Harrow_2009, Chakrabortyinproceedings, Gily_n_2019}. These algorithms require large depth circuits and remain beyond the reach of near and intermediate term devices. One of the main challenges for bringing quantum optimization closer to reality is finding ways to redesign quantum linear systems solvers, and quantum linear algebra more generally, with reduced hardware requirements. In contrast, quantum algorithms for machine learning require full-scale quantum computers and also impose additional requirements (such as the availability of quantum random access memory (QRAM), which is a quantum analogue of classical random access memory (RAM) and could be difficult to construct) \cite{bouland2020prospects}.\input{Figures/Methodologies} 

\subsubsection{Quantum Algorithms}\label{subsubsec:Quantum_Algorithms}

\autoref{fig:Quantum_Algorithm} shows the breakdown of the quantum algorithm layer. The application in the field of Monte Carlo simulation, namely the estimation of risk measures or the pricing of derivatives most often use searching and counting algorithms. However, there is one paper using a quantum algorithm for solving linear equation systems. Most applications related to optimization or portfolio optimization are based on approximate optimization. This is explained by the fact that quantum annealing is mainly used for optimization problems and uses Ising models. However, there are two publications each that use a quantum algorithm to solver linear equation systems and quantum optimization algorithm to find solutions to find optimal portfolios. Note that the related algorithms discussed have been covered in detail in \autoref{subsubsec:Met_Quantum_Algorithms}. \input{Figures/Quantum_Algorithm}

\subsubsection{Hardware}\label{subsubsec:Hardware}

Four of the use cases and their implementations relate to quantum annealer devices and the other nine to gate-based quantum computers, as shown in \autoref{fig:Hardware}. The quantum annealer devices are mainly related to implementations for optimization-related use cases and the Ising model as quantum algorithm. This is not surprising, since quantum annealers are particularly well-suited for optimization problems. Gate-based quantum computers, on the other hand, cover both the optimization and Monte Carlo use cases and are mostly based on QPE or QAE algorithms.\input{Figures/Hardware}

\subsection{Portfolio optimization}\label{subsec:Portfolio_optimization}

In the following, we will explain in more detail the results of the five papers describing portfolio optimization with a quantum computer. In particular, we will explain the domains of the considered use cases and summaries the results.

All papers examine portfolio optimization under classical \citeauthor{Markowitz_1952} portfolio theory \cite{Markowitz_1952}. The problem of identifying the portfolio that maximizes return for a given level of risk (formulated as the standard deviation of returns) is a central task of classical portfolio theory. 
The three papers by \citeauthor{phillipson2020portfolio} \cite{phillipson2020portfolio}, \citeauthor{venturelli2019reverse} \cite{venturelli2019reverse} and \citeauthor{rebentrost2018quantum} \cite{rebentrost2018quantum} discuss portfolio optimization without constraints, which is basically a quadratic optimization problem. Note that constraints are, for example, no short selling or budget constraints.  The main difference between the approach of \citeauthor{rebentrost2018quantum} \cite{rebentrost2018quantum}, \citeauthor{phillipson2020portfolio} \cite{phillipson2020portfolio} and \citeauthor{venturelli2019reverse} \cite{venturelli2019reverse} is that the former allow any fraction of an asset to be held in a portfolio and is intended for gate-based quantum computers, while the other implementations only allow assets to either be held in the portfolio or not and is intended for quantum annealers. 

\citeauthor{rebentrost2018quantum} \cite{rebentrost2018quantum} use the HHL algorithm and its variations to describe the theoretical implementation to identify the minimum risk portfolio for a given desired return. Because there are no inequality constraints, the problem becomes a linear least squares problem for which there is a closed form solution. The approach of \citeauthor{rebentrost2018quantum} \cite{rebentrost2018quantum} is intended for implementation on a gate-based quantum computer. The algorithm can in principle be efficient in the sense that it has a running time proportional to the number of qubits involved in the computations, i.e. the running time is $poly(log(N))$ versus $poly(N)$ with classical algorithms. The hardware requirements are $O(N)$ for storing and accessing the data.

\citeauthor{phillipson2020portfolio} \cite{phillipson2020portfolio} and \citeauthor{venturelli2019reverse} \cite{venturelli2019reverse} restrict to a binary optimization problem, i. e. assets can either be included in the portfolio or not. This type of optimization problem is also called Quadratic Unconstrained Binary optimization (QUBO) problem. \citeauthor{phillipson2020portfolio} \cite{phillipson2020portfolio} describe the implementation of the problem on a quantum annealer and its hybrid solver. They try to find the optimal portfolio of two main stock indices, the Nikkei225 and the S\&P500 indices. The results are compared with conventional tools. The results show that for problems of the size of the example used, certain quantum annealers in its current, still limited scope, are already able to approach the performance of the commercial solvers. \citeauthor{venturelli2019reverse} \cite{venturelli2019reverse} create a hybrid algorithm (classical and quantum) and use real financial data statistics as \citeauthor{phillipson2020portfolio} \cite{phillipson2020portfolio} and formulate the case also as a QUBO problem. The optimized reverse annealing protocol turns out to be more than 100 times faster than the corresponding forward quantum annealing (as described by \citeauthor{phillipson2020portfolio} \cite{phillipson2020portfolio}) on average. However, it is not clear what increase in speed could theoretically be achieved compared to a classical approach. Another question that needs to be clarified is whether the implementation of QUBO can be adapted so that assets in a portfolio can also be weighted arbitrarily, which is of more practical importance. 

The papers discussed above describe promising speed-ups, both based on theoretical as well as experimental results. In practice, however, the portfolio optimization problem often comes with constraints. For example, one can add constraints to ensure that the portfolio is diversified by adding restrictions on the amount of investment in certain asset classes. \citeauthor{kerenidis2019quantum} \cite{kerenidis2019quantum} describe the case of positive portfolio optimization (no short selling) and budget constraints. They emphasize that when there are no inequality constraints (as described by \citeauthor{rebentrost2018quantum} \cite{rebentrost2018quantum}) the Markowitz portfolio optimization problem \cite{Markowitz_1952} becomes a linear least squares problem for which there is a closed form solution. However, if there are constraints there is no closed form solution. \citeauthor{kerenidis2019quantum} \cite{kerenidis2019quantum} reduce this problem to Second Order Cone Programs (SOCPs), which is a more general family of optimization problems and is the main contribution of their work. The problem can be further
solved by considering the quantum SOCP interior-point method of \citeauthor{Kerenidis_2021} \cite{Kerenidis_2021}. Theoretical analysis shows that the running time of this algorithm scales better than classical complexity. However, this speed-up depends on problem-dependent parameters.

The last of our five papers relating to portfolio optimization is by \citeauthor{hodson2019portfolio} \cite{hodson2019portfolio} and deals with portfolio rebalancing which is of special relevance for institutional investors. Portfolio rebalancing is the process of maintaining the original risk and return characteristics of a portfolio over time, as defined, for example, by \cite{Tokat52}. \citeauthor{hodson2019portfolio} \cite{hodson2019portfolio} define portfolio rebalancing as a periodic asset management process in which traders maintain the net value of an institutional portfolio. It is unclear why they refer to the net value of the portfolio rather than the risk and return characteristics. However, as \citeauthor{hodson2019portfolio} \cite{hodson2019portfolio} write, rebalancing can be achieved through a combination of financial instruments, including long and short positions in assets and their derivatives. They also write that because of the computational load of rebalancing, it is typically done overnight (when markets are closed). However, if it can be calculated more quickly, it could enable investors to be more agile in response to changes in markets. They formulated two approaches, one with soft constraints and the Quantum Approximate optimization Algorithm (QAOA), and another with hard constraints and the quantum alternating operator ansatz. The quantum alternating operator ansatz, is the consideration of general parameterized families of unitaries rather than only those corresponding to the time evolution under a fixed local Hamiltonian for a time specified by the parameter \cite{Hadfield2019From}.
The results of the experiments from running the algorithms on an ideal simulator highlight in particular the importance of scaling the input data for the QUAO approach and the performance advantages of the quantum alternating operator approach over it. However, further investigation is needed at several critical points to show the path from the QUAO and/or quantum alternating operator ansatz in this use case to a potential quantum advantage. In particular, the potential, i. e. theoretical speed-up of these approaches compared to a classical approach is not yet clear.

Finally, it is worth mentioning the index tracking problem, which is a variant of the portfolio optimization problem. The index tracking problem aims to select a few assets so that the resulting portfolio's return is as close as possible to a given reference index. For instance, \citeauthor{fernandez2021hybrid} \cite{fernandez2021hybrid}\footnote{This paper is not included in our sample because the algorithm is not yet implemented on quantum hardware.} discuss a hybrid approach to solve optimization problems with constraints. They describe the application of quantum optimization algorithms to index tracking. They numerically simulate the application of quantum optimization algorithms such as QAOA and VQE to this problem, and compare the performance of different quantum variational optimization algorithms in their pruning method.

In summary, the relevance of faster portfolio optimization is confirmed by all authors. There are several research directions and algorithms within portfolio optimization. However, more research is needed to find a clear benefit for the financial industry. 

\subsection{Transaction Settlement}\label{subsec:Transaction_Setllement}

One use case that is closely related to combinatorial optimization is transaction settlement and described by \citeauthor{braine2021quantum} \cite{braine2021quantum}. They focus on securities settlement in the capital markets. By transaction settlement, they refer to the process by which securities (tradable financial assets such as shares, bonds and derivatives) are delivered in exchange for payment. This exchange between parties can be facilitated by a clearing house, which also mitigates counterparty risk. The clearing house's goal is usually to clear as many transactions as possible or to maximise the total value of cleared transactions, but this is a difficult optimization problem due to a combination of legal frameworks and additional optionalities.

\citeauthor{braine2021quantum} \cite{braine2021quantum} emphasise that transaction settlement is a binary quadratic optimizations problem. As discussed in \autoref{subsec:Portfolio_optimization}, algorithms such as the Variational Quantum Eigensolver (VQE) and the Quantum Approximate optimization Algorithm (QAOA) exist for solving quadratic \textit{unconstrained} binary optimization problems (QUBO) with quantum computers. However, many relevant problems in practice are Mixed Binary optimization (MBO) problems, with discrete and continuous variables or with constraints (e. g. inequality constraints) that cannot be modelled as part of a QUBO problem. Since the transaction settlement problem also has corresponding constraints, it also belongs to MBO problems.

The main contribution of \citeauthor{braine2021quantum} \cite{braine2021quantum} is the introduction of an approach to extend the existing quantum methods (VQE and QAOA) to more general Mixed Binary Optimization (MBO) problem classes. They use a hybrid quantum/classical heuristics to address MBO problems which allows one to model optimization with inequality constraints with a quantum computer. Furthermore, they apply this approach to the transaction settlement problem and demonstrate the performance of the corresponding algorithm on concrete instances using cloud-based quantum hardware. Their algorithm presents only part of the transaction settlement problem. However, the algorithm lays the basis for further development and also allows for the use of problem-specific variational forms, such as those proposed by QAOA, which may further improve the performance.

\subsection{Derivative Pricing}\label{subsec:Derivative_Pricing}

The price of derivatives depends on the price of a single asset or multiple assets (e.g. stocks) and other parameters (e.g. interest rate curve). As elaborated by \citeauthor{martin2021toward} \cite{martin2021toward}, due to the stochastic nature of the underlying parameters to which derivatives relate, calculating their fair value can be a difficult task. While analytical models exist for the simplest types of options, the simplifying assumptions about market dynamics required for the models to provide closed-form solutions often limit their applicability. Therefore, in most cases, numerical methods must be used for option pricing, with Monte Carlo being one of the most popular methods due to its flexibility and ability to handle stochastic parameters generically \cite{martin2021toward}. 

In this context, it is important to discuss the difference between path-independent and path-dependent derivatives, which is also highlighted by \citeauthor{Stamatopoulos_2020}. \cite{Stamatopoulos_2020}. Path-independent derivatives have a payoff function, i.e. the function that defines the payoff structure of a financial derivatives at the exercise date that depends on an underlying asset at a single point in time. Therefore, the price of the asset up to the exercise date of the derivative is irrelevant for the price of it. In contrast, the payoff structure of a path-dependent derivative depends on the development of the asset's price and its history up to the exercise date. Derivatives that are path-independent and relate to a single asset are the easiest to value, and in most cases the numerical calculation is straightforward and would, according to \citeauthor{Stamatopoulos_2020} \cite{Stamatopoulos_2020}, probably not benefit from the use of a quantum computer. Path-independent derivatives, which relate to multiple assets, are only slightly more difficult to value because the probability distributions must take into account correlations between assets, but they can usually also be valued quite efficiently on classical computers. Path-dependent derivatives, on the other hand, are much more difficult to price than path-independent derivatives because they require an often computationally expensive payoff calculation at multiple points in time on each path, so minimizing the number of paths required for this step would lead to a significant advantage in the pricing process. It is in this last case that the greatest impact of quantum computing becomes apparent according to \citeauthor{Stamatopoulos_2020} \cite{Stamatopoulos_2020}.

\citeauthor{Rebentrost_2018} \cite{Rebentrost_2018} discuss the Black-Scholes-Merton model \cite{Black1973The, Merton1973The}, which considers the pricing of options, which are a fundamental subset of financial derivatives. They show how the relevant probability distributions can be prepared on a gate-based quantum computer in quantum superposition, the payoff functions can be implemented via quantum circuits, and the price of options can be extracted via quantum measurements. Their most important contribution is the use of the amplitude estimation (AE) algorithm, which can theoretically achieve a quadratic speed-up in the number of steps required to estimate the corresponding price compared to the Monte Carlo approach. In other words, a quadratic speed-up in the number of samples required to estimate the price of the derivative to a given error: If the desired error is $\epsilon$, then classical methods show $1/\epsilon$ dependence in the number of samples, while the quantum algorithm shows $1/\epsilon^{2}$ dependence. Another promising point about this approach is that it can, in principle, be applied to any derivative whose payoff is a function that can be efficiently decomposed into elementary arithmetic operations within a quantum circuit.

\citeauthor{Stamatopoulos_2020} \cite{Stamatopoulos_2020} extend the approach of \citeauthor{Rebentrost_2018} \cite{Rebentrost_2018}, i.e. of using AE algorithms to price options, and place a strong emphasis on implementing the algorithms in a gate-based quantum computer. In detail, they show the implementation of the quantum circuits required to generate the input states and operators needed by AE to price the different option types, and apply a simple error mitigation scheme. Furthermore, they investigate the performance of their approach using a quantum device. The results are affected by readout errors and errors that occur during the execution of the circuits. However, the accuracy of the calculated option price was dramatically increased by the simple error mitigation scheme: it reduced the error averaged over the initial spot price from 62\% to 21\%. They came to the conclusion that much larger quantum hardware capable of running deeper quantum circuits with more qubits than currently available quantum computers is needed to evaluate typical portfolios in the financial industry. Moreover, the results suggest that developments in quantum hardware, such as error corrections, are needed to perform practical-relevant experiments and achieve speed-up compared to current applied methods.

As mentioned above, there are many derivative financial instruments for which there is no general analytical solution and which are solved in practice with numerical simulations. For these, however, there is a trade-off between the number of stochastic factors considered and the computational time to run a numerical simulation. The multi-factor Heath-Jarrow-Morton (HJM) model \cite{heath1992bond} is widely used in finance for the valuation of interest rate derivatives. However, the main problem in building an HJM model to describe the development of the yield curve is the large number of degrees of freedom. One possible selection of factors is based on principal component analysis (PCA). Besides, there is also factor analysis or selection based on economic intuition \cite{sabelli2015multicurve}. \citeauthor{martin2021toward} \cite{martin2021toward} show how to apply quantum principal component analysis (qPCA) to reduce the number of stochastic factors to find the fair price of a interest rate derivative. Moreover, they implement this algorithm in a quantum computer. However, the experiments exhausted the computational power offered by the current quantum processors in terms of gate fidelity, connectivity and number of qubits. Nevertheless, their work is a first step towards the development of a general quantum algorithm to fully simulate the HJM model for pricing interest rate derivatives on quantum computers.

In summary, the three contributions already describe in detail the possible use of quantum computers for the pricing of different types of derivatives. Furthermore, by using the AE algorithm, a theoretical quadratic speed-up is shown compared to classical Monte Carlo simulations. Moreover, simple experiments have already been implemented on real quantum computer devices. However, these experiments also show that extensive hardware developments are still necessary for a practical application of quantum computer for derivative pricing. However, it was also shown that improvements can be achieved with error corrections and thus a practical application on near-term noisy intermediate-scale quantum computers could exist.

\subsection{Estimating Risk Measure(s)}\label{subsec:Estimating_Risk_Measures}

Two papers discuss how risk measures can be estimated with quantum computers. In this context, it is important to note that there are many different risk measures for different types of risks. However, one of the most popular methods for quantitatively assessing the risk of portfolios consisting of assets such as stocks, credits, or fixed income securities is the Value at Risk (VaR). In short, it is a potential loss that will not be exceeded with a certain probability, defined as the confidence level, over a given time period and for a given portfolio of financial assets \cite{bessis2015risk}. While the basic concept of VaR is simple, many complications can arise in practice. One important complication is the non-linearity in the portfolio payoff structure caused, for example, by derivatives such as options \cite{ammann2001var}. The Monte Carlo method is one of the methods to overcome this problem \cite{Glasserman2000efficient}. 

\citeauthor{WE2019} \cite{WE2019} present in their paper a quantum algorithm that analyzes risk measures such as VaR. Specifically, they show how to use quantum AE to evaluate the risk measures VaR and Conditional VaR (CVaR) on a gate-based quantum computer. \citeauthor{WE2019} \cite{WE2019} has implemented also a small example of two assets on a real quantum-computer device. However, this example is of very low complexity and no nonlinear payoff structures were considered. For more industry relevant examples and to achieve a practical relevant speed-up, more qubits are needed and the errors of the current hardware must be reduced. In addition, they make another important point, namely that Monte Carlo simulations can be massively parallelised in practice, which pushes the bound for quantum advantage even higher.

In a follow-up paper, \citeauthor{egger2019credit} \cite{egger2019credit} build on this work and use a very similar approach, and thus also show a theoretical quadratic speed-up to estimate the difference between the VaR and the expected loss value. This risk measure is also known as the economic capital requirement. They also estimate the circuit depth of applications of this method for practically relevant use cases. The requirements for such use cases are shown to be well beyond the capabilities of current gate-based quantum hardware. However, as already discussed above by other authors, they concluded that research in algorithms can help reduce the number of qubits as well as the circuit depth so that quantum computers will find applications in finance sooner.

\subsection{Predicting Financial Crashes}\label{subsec:Predicting_Financial_Crashes}

Two papers demonstrate the prediction of financial crises using a quantum computer. However, both papers address the same approach. In the paper by \citeauthor{Orus2019} \cite{Orus2019} the basis is elaborated. The later paper by \citeauthor{ding2019prediction} \cite{ding2019prediction} then shows the implementation of this method on a quantum annealer and corresponding results.

The financial network (banks, companies, individuals) holds a set of assets as well as some of the other institutions in the network. The question is what change in the value of the assets could cause a massive drop in the market value of the institutions, i.e., a financial crash. According to \citeauthor{Orus2019} \cite{Orus2019}, this type of problem is extremely difficult to solve even for simple models, making it difficult and almost unsolvable for classical computers. Nevertheless, this problem is of great importance from a regulatory and macroeconomic perspective. \citeauthor{Orus2019} \cite{Orus2019} propose a nonlinear network model for financial markets to predict such financial crashes. Their main contribution is the efficient construction of a QUBO formula (see also \autoref{subsec:Portfolio_optimization}) representing the problem. They also show how to find a solution using a quantum annealer.

\citeauthor{ding2019prediction} \cite{ding2019prediction} then implemented this proposed algorithm on a quantum annealer. According to the authors, this work can be considered the first proof that quantum computation can be used to study quantitative finance and help institutions anticipate risks. However, a next-generation quantum processor is needed to implement larger networks and therefore relevant computations for the financial industry. This result could be sooner achieved by developing a problem-specific quantum annealer.

Although the approach sounds promising, the prediction of financial crashes is generally considered very difficult (see also \autoref{subsubsec:Met_Use_Cases}). Therefore, besides the improvement of quantum hardware, it has to be proven whether the chosen approach of the network model can indeed predict financial crashes with high confidence. This could be shown, for example, with back-testing using historical data.

%% file: Figures/Zwicky-Box.tex
\begin{figure*}[ht]\small
\begin{tabularx}{\textwidth}{|m{1.8cm}|C|C|C|C|C|C|}
\hline
\parbox[b][2.5cm][c]{\linewidth}{\textbf{Use Case}} & \markZwicky(PO){\parbox{\linewidth}{\centering Portfolio optimisation}} & \markZwicky(TS){\parbox{\linewidth}{\centering Transaction settlement}} & \markZwicky(PFC){\parbox{\linewidth}{\centering Predicting financial crashes}} & \markZwicky(ERM){\parbox{\linewidth}{\centering Estimating risk measure(s)}} & \markZwicky(DP){Derivative pricing} \\
\hline
\end{tabularx}\\[\vcorrect]
\begin{tabularx}{\textwidth}{|m{1.8cm}|C|C|C|}
\parbox[b][2.5cm][c]{\linewidth}{\textbf{Methodology}} & \markZwicky(O){Optimisation} & \markZwicky(MC){Monte Carlo} & \markZwicky(ML){Machine learning} \\
\hline
\end{tabularx}\\[\vcorrect]
\begin{tabularx}{\textwidth}{|m{1.8cm}|C|C|C|C|C|C|}
\parbox[b][2.5cm][c]{\linewidth}{\textbf{Quantum\\ algorithm}} & 

\markZwicky(AO){\parbox{\linewidth}{\centering Approximate Optimization (VQE QAOA, QUBO)}} & \markZwicky(OP){\parbox{\linewidth}{\centering Optimization (QIPM)}} & \markZwicky(SLES){\parbox{\linewidth}{\centering Solving linear equation systems (HHL, qPCA)}} & \markZwicky(SC){\parbox{\linewidth}{\centering Searching and counting (QAE, QPA, QPE)}}  \\
\hline
\end{tabularx}\\[\vcorrect]
\begin{tabularx}{\textwidth}{|m{1.8cm}|C|C|}
\parbox[b][2.5cm][c]{\linewidth}{\textbf{Hardware}} & \markZwicky(QA){Quantum annealer} &  \markZwicky(GB){Gate-based quantum computer}  \\
\hline
\end{tabularx}
\tikzZwicky[blue](PO)(O)(SLES)(GB) 
\tikzZwicky[blue](PO)(O)(AO)(QA) 
\tikzZwicky[blue](TS)(O)(AO)(GB) 
\tikzZwicky[blue](PO)(O)(AO)(QA) 
\tikzZwicky[blue](PFC)(O)(AO)(QA) 
\tikzZwicky[blue](PO)(O)(AO)(QA) 
\tikzZwicky[blue](PO)(O)(OP)(QA) 
\tikzZwicky[blue](PFC)(O)(AO)(QA) 
\tikzZwicky[blue](DP)(MC)(SC)(GB) 
\tikzZwicky[blue](DP)(MC)(SLES)(GB) 
\tikzZwicky[blue](ERM)(MC)(SC)(GB) 
\tikzZwicky[blue](ERM)(MC)(SC)(GB) 
\tikzZwicky[blue](DP)(MC)(SC)(GB) 

\caption{Morphological box representing paper selection}
\label{fig:morphologicalbox}
\end{figure*}






%% file: Figures/Use_Cases.tex
\begin{figure}[h]
	\centering
	\begin{tikzpicture}
	\draw (0cm,0cm) -- (5cm,0cm);  
	\draw (0cm,0cm) -- (0cm,-0.1cm);  
	\draw (5cm,0cm) -- (5cm,-0.1cm) node [below] {\makecell[c]{}};  
	
	\draw (-0.1cm,0cm) -- (-0.1cm,3cm);  
	\draw (-0.1cm,0cm) -- (-0.2cm,0cm);  
	\draw (-0.1cm,3cm) -- (-0.2cm,3cm) node [left] {Number};  
	
	\foreach \x/\m in {
		0.5/0,
		1.0/4,
		1.5/8,
		2.0/12,
		2.5/16
	}  
	\draw[gray!50, text=black] (-0.2 cm,\x cm) -- (5 cm,\x cm) 
	node at (-0.5 cm,\x cm) {\m};  

	\foreach \x/\y/\country in {
		0/5/{Portfolio optimisation},  
		1/1/Transaction settlement,
		2/2/Predict. financ. crashes,
		3/2/Estimat. risk measure(s),
		4/3/Derivative pricing
	}
	{
		\draw[draw=black!60, fill=black!15] (\x*1 cm +0.35 cm,0cm) rectangle (0.7cm+\x*1 cm,\y*0.1666 cm) 
		node at (0.55cm + \x*1 cm,\y*0.1666 cm + 0.3cm) {\y}; 
		\node[rotate=45, left] at (0.6 cm +\x*1 cm,-0.1cm) {\country}; 
	};
	\end{tikzpicture}
	\caption{Breakdown of quantum computing applications for finance according to the "use case" layer.}
	\label{fig:Use_Cases}
\end{figure}

%% file: Figures/Methodologies.tex
\begin{figure}[h]
	\centering
	\begin{tikzpicture}
	\draw (0cm,0cm) -- (3cm,0cm);  
	\draw (0cm,0cm) -- (0cm,-0.1cm);  
	\draw (3cm,0cm) -- (3cm,-0.1cm) node [below] {\makecell[c]{}};  
	
	\draw (-0.1cm,0cm) -- (-0.1cm,3cm);  
	\draw (-0.1cm,0cm) -- (-0.2cm,0cm);  
	\draw (-0.1cm,3cm) -- (-0.2cm,3cm) node [left] {Number};  
	
	\foreach \x/\m in {
		0.5/0,
		1.0/4,
		1.5/8,
		2.0/12,
		2.5/16
	}  
	\draw[gray!50, text=black] (-0.2 cm,\x cm) -- (3 cm,\x cm) 
	node at (-0.5 cm,\x cm) {\m};  

	\foreach \x/\y/\country in {
		0/8/{Optimisation},  
		1/5/Monte Carlo,
		2/0/Machine learning
	}
	{
		\draw[draw=black!60, fill=black!15] (\x*1 cm +0.35 cm,0cm) rectangle (0.7cm+\x*1 cm,\y*0.1666 cm) 
		node at (0.55cm + \x*1 cm,\y*0.1666 cm + 0.3cm) {\y}; 
		\node[rotate=45, left] at (0.6 cm +\x*1 cm,-0.1cm) {\country}; 
	};
	\end{tikzpicture}
	\caption{Breakdown of quantum computing applications for finance according to the "use case" layer.}
	\label{fig:Methodologies}
\end{figure}

%% file: Figures/Quantum_Algorithm.tex
\begin{figure}[h]
	\centering
	\begin{tikzpicture}
	\draw (-0.2cm,0cm) -- (5.2cm,0cm);  
	\draw (-0.2cm,0cm) -- (-0.2cm,-0.1cm);  
	\draw (5.2cm,0cm) -- (5.2cm,-0.1cm) node [below] {\makecell[c]{}};  
	
	\draw (-0.3cm,0cm) -- (-0.3cm,3cm);  
	\draw (-0.3cm,0cm) -- (-0.4cm,0cm);  
	\draw (-0.3cm,3cm) -- (-0.4cm,3cm) node [left] {Number};  
	
	\foreach \x/\m in {
		0.5/0,
		1.0/4,
		1.5/8,
		2.0/12,
		2.5/16
	}  
	\draw[gray!50, text=black] (-0.4 cm,\x cm) -- (5.2 cm,\x cm) 
	node at (-0.7 cm,\x cm) {\m};  

	\foreach \x/\y/\country in {
		0/4/{Searching and counting},  
		1/2/{Solving linear equation systems},
		2/1/{Optimization},
		3/6/{Approximate optimization}
	}
	{
		\draw[draw=black!60, fill=black!15] (\x*1.4 cm +0.35 cm,0cm) rectangle (0.7cm+\x*1.4 cm,\y*0.1666 cm) 
		node at (0.55cm + \x*1.4 cm,\y*0.1666 cm + 0.3cm) {\y}; 
		
	};
		\node[rotate=45, align=right, xshift=-2.1cm] at (0.5cm,0cm) {Searching and counting\\(QAE, QPA, QPE)};
		\node[rotate=45, align=right, xshift=-2.1cm] at (1.9cm,0cm) {Solving linear equation\\systems (HHL, qPCA)};
		\node[rotate=45, align=right, xshift=-1.4cm] at (3.3cm,0cm) {Optimization\\(QIPM)};
		\node[rotate=45, align=right, xshift=-2.25cm] at (4.7cm,0cm) {Approximate optimization\\(VQE, QAOA, QUBO)};
	
	\end{tikzpicture}
	\caption{Breakdown of quantum computing applications for finance according to the "quantum algorithm" layer.}
	\label{fig:Quantum_Algorithm}
\end{figure}
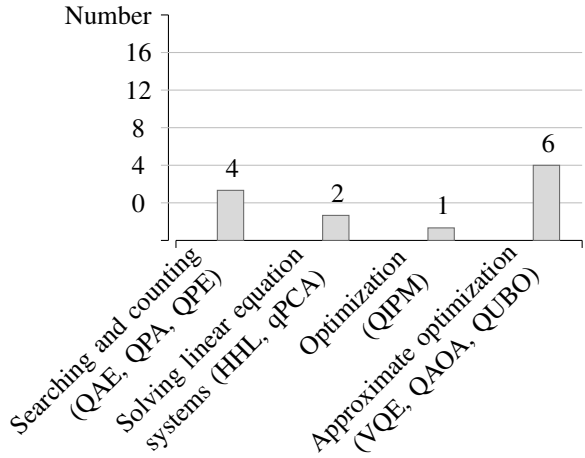

%% file: Figures/Hardware.tex
\begin{figure}[h]
	\centering
	\begin{tikzpicture}
	\draw (0cm,0cm) -- (2cm,0cm);  
	\draw (0cm,0cm) -- (0cm,-0.1cm);  
	\draw (2cm,0cm) -- (2cm,-0.1cm) node [below] {\makecell[c]{}};  
	
	\draw (-0.1cm,0cm) -- (-0.1cm,3cm);  
	\draw (-0.1cm,0cm) -- (-0.2cm,0cm);  
	\draw (-0.1cm,3cm) -- (-0.2cm,3cm) node [left] {Number};  
	
	\foreach \x/\m in {
		0.5/0,
		1.0/4,
		1.5/8,
		2.0/12,
		2.5/16
	}  
	\draw[gray!50, text=black] (-0.2 cm,\x cm) -- (2 cm,\x cm) 
	node at (-0.5 cm,\x cm) {\m};  

	\foreach \x/\y/\country in {
		0/9/Gate-based QC,  
		1/4/{Quantum annealer}
	}
	{
		\draw[draw=black!60, fill=black!15] (\x*1 cm +0.35 cm,0cm) rectangle (0.7cm+\x*1 cm,\y*0.1666 cm) 
		node at (0.55cm + \x*1 cm,\y*0.1666 cm + 0.3cm) {\y}; 
		\node[rotate=45, left] at (0.6 cm +\x*1 cm,-0.1cm) {\country}; 
	};
	\end{tikzpicture}
	\caption{Breakdown of quantum computing applications for finance according to the "quantum hardware" layer.}
	\label{fig:Hardware}
\end{figure}

%% file: Chapters/05_Conclusion.tex
\section{Conclusion}\label{sec:Conclusion}

In this paper, we discussed the current state of quantum computing in the financial sector and systematically analyzed potential use cases for quantum computing in finance. For this purpose, we conducted an extensive literature search and designed a multi-layered framework to enable a structured analysis of the available literature and the use cases described. The corresponding use cases focus on the areas of optimization and Monte Carlo simulation. The first area in particular has significant potential in the context of investment portfolio construction. The use of quantum computing could be beneficial here in that optimal portfolios can be found more quickly. In the area of Monte Carlo methods, applications are described primarily with regard to the pricing of derivatives and the estimation of risk measures such as VaR or credit default risk. In particular, faster or more accurate pricing of derivatives could greatly improve the liquidity of financial markets, while more accurate estimation of risk measures could lead to more stable financial institutions. However, no paper has been found that describes a use case for quantum-based machine learning in finance. This is somewhat surprising as machine learning already has many applications in traditional finance. 

There are already many different use cases for quantum computing in different areas of finance, but they differ in many aspects. Some of them have already been implemented on a real quantum device. Nevertheless, it needs to be mentioned that breakthroughs in the development of quantum computing hardware are necessary to realize the full potential of quantum-based algorithms for finance. For example, more qubits are needed and the errors of current hardware need to be reduced. The development of quantum computers is highly complex which is one of the reasons why little literature exists on the impact on the financial industry. However, existing publications point to the increase in investment profits, reduction in capital requirements, creation of new investment opportunities, and improvement in risk management and compliance. To unlock these potentials, quantum computing must be operated economically and cost effectively.

\input{Figures/Table_Papers}

%% file: Figures/Table_Papers.tex
\onecolumn

\begin{table}[H]\footnotesize
\caption{Overview of proposed quantum algorithm for use cases in finance.}
\label{tab:Table}
\centering
\begin{tabular}{p{1.75cm}p{0.75cm}p{3.3cm}p{3.5cm}p{1.75cm}p{2.4cm}p{2cm}}

\toprule
\textbf{Author(s)}&\textbf{Year}&\textbf{Title}&\textbf{Use Case}&\textbf{Methodology}&\textbf{Quantum algorithm}&\textbf{Hardware}\\\midrule

\citeauthor{rebentrost2018quantum}&\citeyear{rebentrost2018quantum}&\citetitle{rebentrost2018quantum}&Determine the optimal portfolio more quickly/accurately.&Optimisation&Solving linear equation systems (HHL)&Gate-based quantum computer\\\hdashline
\citeauthor{venturelli2019reverse}&\citeyear{venturelli2019reverse}&\citetitle{venturelli2019reverse}&Determine the optimal portfolio more quickly/accurately.&Optimisation&Approximate optimization (QUBO)&Quantum annealer\\\hdashline
\citeauthor{braine2021quantum}&\citeyear{braine2021quantum}&\citetitle{braine2021quantum}&Settle as many transactions as possible and/or maximise the total value of the settled transactions.&Optimisation&Approximate optimization (VQE and QAOA)&Gate-based quantum computer\\\hdashline

\citeauthor{phillipson2020portfolio}&\citeyear{phillipson2020portfolio}&\citetitle{phillipson2020portfolio}&Determine the optimal portfolio more quickly/accurately.&Optimisation&Approximate optimization (QUBO)&Quantum annealer\\\hdashline
\citeauthor{ding2019prediction}&\citeyear{ding2019prediction}&\citetitle{ding2019prediction}&Prediction of financial crashes in a complex financial network.&Optimisation&Approximate optimization (QUBO adapted)&Quantum annealer\\\hdashline

\citeauthor{hodson2019portfolio}&\citeyear{hodson2019portfolio}&\citetitle{hodson2019portfolio}&More quickly/accurately discrete portfolio optimisation under constraints.&Optimisation&Approximate optimization (QAOA)&Gate-based quantum computer\\\hdashline
\citeauthor{kerenidis2019quantum}&\citeyear{kerenidis2019quantum}&\citetitle{kerenidis2019quantum}&More quickly/accurately portfolio optimisation under constraints.&Optimisation&Optimization (QIPM for SOCPs)&Gate-based quantum computer\\\hdashline
\citeauthor{Orus2019}&\citeyear{Orus2019}&\citetitle{Orus2019}&Forecasting financial crashes with quantum computing.&Optimisation&Approximate optimization (QUBO)&Quantum annealer\\\hdashline

\citeauthor{Rebentrost_2018}&\citeyear{Rebentrost_2018}&\citetitle{Rebentrost_2018}&Price derivatives more quickly/accurately&Monte Carlo&Searching and counting (QAE)&Gate-based quantum computer\\\hdashline
\citeauthor{martin2021toward}&\citeyear{martin2021toward}&\citetitle{martin2021toward}&Pricing interest-rate financial derivatives with the Heath-Jarrow-Morton model more quickly&Monte Carlo&Optimization (qPCA)&Gate-based quantum computer\\\hdashline
\citeauthor{egger2019credit}&\citeyear{egger2019credit}&\citetitle{egger2019credit}&Estimate credit risk more efficiently.&Monte Carlo&Searching and counting (QAE)&Gate-based quantum computer\\\hdashline

\citeauthor{WE2019}&\citeyear{WE2019}&\citetitle{WE2019}&Evaluate risk measures such as Value at Risk and Conditional Value at Risk more quickly/accurately&Monte Carlo&Searching and counting (QAE)&Gate-based quantum computer\\\hdashline

\citeauthor{Stamatopoulos_2020}&\citeyear{Stamatopoulos_2020}&\citetitle{Stamatopoulos_2020}&Price options such as vanilla options, multi-asset options, barrier options and path-dependent options more quickly/accurately&Monte Carlo&Searching and counting (QAE)&Gate-based quantum computer\\\bottomrule
\end{tabular}
\end{table}

\twocolumn